\documentclass[12pt,tightenlines,eqsecnum,floats,aps,amsmath,amssymb,nofootinbib,prd,showpacs]{revtex4}

\usepackage{setspace}
\usepackage{amsmath,amssymb,amsfonts,amsthm}
\usepackage{graphicx}
\usepackage{enumerate} 
\usepackage{colordvi} 

\def\be{\begin{equation}}
\def\ee{\end{equation}}
\def\ba{\begin{eqnarray}}
\def\ea{\end{eqnarray}}
\def\bi{\begin{itemize}}
\def\ei{\end{itemize}}
\def\bnum{\begin{enumerate}}
\def\enum{\end{enumerate}}

\def\b{\bar}

\def\f{\frac}

\def\R{\mathbb{R}}

\def\ub{\underbar}

\def\ep{\epsilon}

\def\utw#1{\rlap{\lower1ex\hbox{$\sim$}}#1{}} 

\usepackage{enumerate}

\usepackage{colordvi}
\newcounter{mnotecount}[section]

\newcommand{\comment}[1]{}
\begin{document}

\title{Hamiltonian General Relativity and the\\ Belinskii, Khalatnikov,
Lifshitz Conjecture}

\author{Abhay Ashtekar}\email{ashtekar@gravity.psu.edu}
\author{Adam Henderson}\email{adh195@psu.edu}
\author{David Sloan}\email{sloan@gravity.psu.edu}
 \affiliation{Institute for Gravitation and the Cosmos \& Physics
 Department,
 Penn State, University Park, PA 16802-6300, U.S.A.}

\begin{abstract}
The Belinkskii, Khalatnikov and Lifshitz conjecture says that as one
approaches space-like singularities in general relativity, `time
derivatives dominate over spatial derivatives' so that the dynamics
at any spatial point is well captured by an ordinary differential
equation. By now considerable evidence has accumulated in favor of
these ideas. Starting with a Hamiltonian framework, we provide a
formulation of this conjecture in terms of variables that are
tailored to non-perturbative quantization. Our formulation serves as
a first step in the analysis of the fate of generic space-like
singularities in loop quantum gravity.
\end{abstract}


\maketitle

\section{Introduction}
\label{s1}

In 1970, Belinskii, Khalatnikov and Lifshitz (BKL) made a conjecture
which, if true, would shed considerable light on the nature of
space-like singularities in general relativity \cite{bkl}. They
suggested that as one approaches these singularities, time
derivatives would dominate over spatial derivatives, implying that
the asymptotic dynamics would be well described by an ordinary
differential equation. At first this claim seems astonishingly
strong. However, over the years considerable analytic and numerical
evidence has accumulated in its favor (see, e.g., \cite{bb,ar,dg1}).
Together, these results suggest the following scenario: i) Geometry
at any spatial point is well described by the Bianchi I metric for
long stretches of time; ii) From time to time the parameters $p_i$
characterizing the Bianchi I metric undergo a (Bianchi II)
transition, rapidly settling to new values $p'_i$; and, iii) The
spatial gradients \emph{can} grow but they do so on `small sets'
where spikes develop.%
\footnote{As is common in the recent literature, we use the
convention that the singularity occurs at $t= -\infty$. So in
general there are infinitely many transitions till one reaches the
singularity. However, general relativity cannot be trusted once the
curvature acquires Planck scale and until this epoch there are only
a finite number of transitions and spikes at any spatial point.}
In addition, except for a scalar field or a stiff fluid, sources
do not play much of a role in dynamics; `matter doesn't matter'
close to the singularity.

It is tempting to use this scenario as a starting point in the
analysis of what happens to these classical singularities in the
quantum theory. In particular, Garfinkle \cite{dg2} has suggested
that understanding of the quantum behavior of the Bianchi I model
would shed considerable light on the fate of space-like
singularities in quantum gravity. It is now known in loop quantum
cosmology that the Bianchi I singularity is naturally resolved
because of quantum geometry effects \cite{awe2,cv}. This suggests
that there may well be a general result which says that \emph{all}
space-like singularities of the classical theory are naturally
resolved in loop quantum gravity.

However, it is difficult to test this idea using the current
formulations \cite{cu,td} of the BKL conjecture. For, these
formulations are motivated by the theory of partial differential
equations rather than by Hamiltonian or quantum considerations. The
basic variables, for example, are chosen to simplify functional
analysis and numerics. While this is a natural strategy within
classical general relativity, these variables are not well suited
for quantum theory. For example, one often uses the variables
$\Sigma_{ab}, N_{ab}$, introduced by Uggla, van Elst, Wainwright and
Ellis (UEWE), which are obtained by dividing certain geometric
fields by the trace $K$ of the extrinsic curvature \cite{cu}. It is
difficult to write down the corresponding operators in the quantum
theory particularly because of the $K^{-1}$ factors. So, the
question arises: Is there a formulation of the BKL conjecture in
terms of variables that are well suited for non-perturbative
quantization? The answer turns out to be in the affirmative. The
goal of this communication is to present a succinct summary of this
formulation. As we will see, the framework has some features which
make it appealing also in the classical theory.

We will focus on vacuum general relativity because `matter does not
matter'. Details for the scalar field
---which does matter--- have been worked out and key differences
will be summarized in section \ref{s5}.

\section{Convenient Variables}
\label{s2}

We will consider space-times $({}^4\!M, {}^4\!g_{ab})$ with
${}^4\!M= M\times \R$, where $M$ is a compact 3-manifold without
boundary. Let us choose as our gravitational variables pairs
$(E^a_i, K_a^i)$ of fields on $M$, where $a,b,c \ldots$ are tensor
indices and $i,j,k \ldots$ are ${\rm SO(3)}$ internal indices (which
can be freely raised and lowered using the Cartan Killing metric on
${\rm so(3)}$). $E^a_i$ represents an orthonormal triad on $M$ (with
density weight 1) and $K_a^i$ represents extrinsic curvature `on
shell'. These are canonically conjugate on the gravitational phase
space \cite{aa1,jr}: $\{E^a_i (x),\, K_b^j(y)\}\, =
\delta^a_b\,\delta_i^j\, \delta^3(x,y)$.

The density weighted triad $E^a_i$ determines a positive definite
metric $q_{ab}$ on $M$ via $E^a_i E^{bi}= q\, q^{ab}$ where $q$ is
the determinant of $q_{ab}$. The standard extrinsic curvature
$K_{ab}$ is given by $E^{bi} K_{ab} = \sqrt{q}\, K_a^i$. The
scalar, vector and Gauss constraints of vacuum general relativity
are given by
\be S:= -q\,R - 2 E^a_{[i}E^b_{j]}\, K_a^i K_b^j \approx 0, \quad
V_a:= 4\, D_{[a}\,(K_{b]}^iE^b_I) \approx 0, \quad G^k
:=\epsilon_{i}{}^{jk} E^a_{j} K_a{}^i \approx 0\, ,\ee
where $D$ and $R$ denote the derivative operator and the Ricci
scalar of $q_{ab}$. As usual, the Hamiltonian generating dynamics is
just a linear combination of these constraints. The triad $E^a_i$
determines a unique ${\rm SO(3)}$ connection $\Gamma_a^i$ through
$D_a E^b_i + \epsilon_{ij}{}^k \Gamma_a^j E^b_k = 0$. One can define
another {\rm $SO(3)$} connection $A_a^i$ via $A_a^i:= \Gamma_a^i -
\gamma\, K_a^i$ where $\gamma$ is called Barbero-Immirzi parameter.
Loop quantum gravity is based on the canonical pair $(A_a^i\,
E^a_i)$. This pair provides the point of departure for our
formulation of the BKL conjecture.

In the classical analysis we now wish to undertake, it is simpler to
use all three fields, $\Gamma_a^i, K_a^i, E^a_i$, although
$\Gamma_a^i$ is determined by $E^a_i$. The key idea can be then
summarized as follows. The accumulated results to date suggest that
the metric $q_{ab}$ becomes degenerate at the singularity, whence
its determinant $q$ vanishes there. (For example in the Bianchi I
model, in terms of the commonly used proper time $t$, the metric is
given by $ds^2 = -dt^2 + \sum_i t^{2p_i}dx_i^2$ and the singularity
occurs at $t=0$. Since $\sum p_i =1$, we have $q =t^2$.) Therefore
one might expect that fields which are rescaled by appropriate
powers of $q$ would remain well behaved at the singularity.
Similarly, while the covariant spatial derivatives $D_af$ of fields
$f$ may not be sub-dominant compared to time-derivatives,
derivatives $D_i f:= E^a_iD_a f$ are more likely to be, because of
the $\sqrt{q}$ factor in the density weighted triad $E^a_i$. This
strategy is similar to that used in a more common formulation of the
conjecture \cite{bb,dg1,cu} where, as mentioned in section \ref{s1},
one divides geometric fields by the trace $K$ of the extrinsic
curvature which is expected to diverge at the singularity. The
relation between the two strategies is discussed in section
\ref{s3}.

These motivations lead us to introduce the following basic
variables:
\be C_i{}^j:= E^a_i \Gamma_a^j - E^a_k \Gamma_a^k\, \delta_i^j ,
\quad {\rm and} \quad P_i{}^j:= E^a_iK_a^j - E^a_kK_a^k\,
\delta_i^j \,. \ee
It turns out that constraints can be re-expressed \emph{entirely}
in terms of $C_{ij}, P_{ij}$ and their $D_i$ derivatives:
\ba \label{S1} S &:=& 2\ep^{ijk} D_i (C_{jk}) + 4 C_{[ij]}
C^{[ij]} + C_{ij}
C^{ji} - \f{1}{2} C^2 + P_{ij} P^{ji} - \f{1}{2} P^2 \approx 0\\
\label{V1} V_i &:=& - 2D_j P_{i}{}^j - 2 \epsilon_{jkl}
P_i{}^j\, C^{kl} + \epsilon_{ijk} (2P^{jl}C_l{}^k - C P^{jk}) \approx 0\\
\label{G1} G^{k} &:=& \epsilon^{ijk} P_{ji} \approx 0 \, .\ea
Next, let us consider evolution equations, i.e., the Hamiltonian
flow generated by the constraints. For simplicity, in this brief
communication we will set the shift to zero. As in loop quantum
gravity, our lapse $N$ will be a scalar density of weight -1. Then
the time evolution of our basic triplet $(C_{ij},P_{ij}, D_i)$ is
governed by:
\ba \label{Cdot1}\dot{C}_{ij} &= &  \f{1}{2}\,\ep_j{}^{kl} D_k({N}(
 2 P_{li}- \delta_{li} P)) + {N}\, [2 C_{(i|k|} P^k{}_{j)}
+2 C_{[kj]} P^k{}_i - P C_{ij}]\\
\label{Pdot1}\dot{P}_{ij} &=&-\ep_j{}^{kl} D_k ({N} C_{li}) +
\frac{1}{2}\ep_{ij}{}^k D_k ({N} C)- \ep^{klm} D_m ({N} C_{kl}) \delta_{ij} \\
\nonumber & & + 2\ep_{jk}{}^{m} C_{[ik]}\, D_m N +\, (D_i D_j -
\delta_{ij} D^k D_k )\, N \\ \nonumber
& & + N [- 2 C_{(ik)} C^k{}_j + C C_{ij} - 2 C^{[kl]} C_{[kl]}\,\delta_{ij}] \\
\label{Ddot1} \dot{D}_i s_n &=& \f{n}{2}\, [D_i NP]s_n \,-\, N
P_i{}^jD_j s_n \ea
where, in the last equation, $s_n$ is any scalar density of weight
$n$. These equations can be used as follows. On an initial slice,
we construct $(C_{ij}, P_{ij}, D_i)$ from a pair $(E^a_i, K_a^i)$
of canonical variables. But then we can deal exclusively with the
triplet $(C_{ij}, P_{ij}, D_i)$. The pair $(E^a_i, K_a^i)$
satisfies constraints if and only if the triplet satisfies
(\ref{S1})--(\ref{G1}). Given such a triplet, we can evolve it
using (\ref{Cdot1})--(\ref{Ddot1}), \emph{without having to refer
back to the original canonical pair} $(E^a_i, K_a^i)$.

These two sets of equations have some interesting unforeseen
features. First, the basic triplet $(C_{ij}, P_{ij}, D_i)$ has
\emph{only internal indices}: our basic fields are \emph{scalars}
(with density weight 1). It would be of considerable interest to
investigate if this fact provides new insights into the dynamics of
3+1 dimensional gravity \cite{bh}. Second, these equations do not
refer to the triad $E^a_i$. Suppose we begin at an initial time
where $C_{ij}$ is derived from an $E^a_i$. Then these constraint and
evolution equations ensure that $C_{ij}$ is derivable from a triad
at all times. Furthermore, we can easily construct that triad
directly from a solution $(C_{ij}, P_{ij})$ to these equations:
first solve (\ref{Cdot1})--(\ref{Ddot1}) and then simply integrate
the ordinary differential equation
\be \label{Edot} \dot{E}^a_i =  -{N} P_i{}^j E^a_j\,  \ee
at the end. Third, the structure of the constraint and evolution
equations in terms of $(C_{ij}, P_{ij}, D_i)$ is remarkably simple
since only low order polynomials of these variables are involved.
Finally, thanks to our rescaling by $\sqrt{q}$, our basic triplet
$C_{ij}, P_{ij}, D_i$ (as well as $E^a_i$) is expected to have a
well behaved limit at the singularity. (This expectation is borne
out in the Bianchi I and II models.) Note also that our equations
are meaningful even when the triad becomes degenerate. So, strictly
(as in loop quantum gravity \cite{aa1,jr}) we have a generalization
of Einstein's equations. To summarize, we have found variables which
are likely to remain finite at the singularity and rewritten
Einstein's equations as a \emph{closed system of differential
equations} in terms of them. Therefore, this formulation may be
particularly useful for proving global existence and uniqueness
results .

However, in the formulation given above, all variables carry density
weights. Also, although the operator $D_i$  satisfies linearity and
the Leibnitz rule, it is not a standard derivative operator: it has
torsion and, more importantly, it changes the density weight by 1.
These unconventional features may make these equations awkward to
handle particularly for numerical work. In the detailed paper
\cite{ahs2} we will provide an equivalent formulation involving only
functions (with zero density weight) and a more familiar version of
the operator $D_i$.

\section{The conjecture}
\label{s3}

The key step in any formulation of the BKL conjecture is to specify
the basic variables, what one means by their `spatial derivatives'
(which are to be sub-dominant), and `time derivatives' (which are to
dominate). We will consider any smooth foliation $M_t$ of an
appropriate portion of ${}^4\!M$ such that the space-like
singularity of interest constitutes a (limiting) leaf. The time
function $t$ labeling our spatial slices is intertwined with the
choice of lapse. We will assume that the density weighted lapse $N$
admits a smooth limit as one approaches the singularity. Now, we are
led by the intuition that the spatial metric $q_{ab}(t)$ becomes
degenerate at the singularity.  This implies that the lapse function
$\bar{N}$ (with density weight zero), given by $\bar{N} :=
\sqrt{q}N$, goes to zero, i.e., that the singularity is at
$t=-\infty$.

In this set up, our basic variables will be $(C_{ij}, P_{ij})$ and
the lapse $N$. By \emph{time derivatives}, we will mean their Lie
derivatives along the vector field $t^a := \bar{N} n^a$ where $n^a$
is the unit normal to the foliation $M_t$. By \emph{spatial
derivatives} we will mean their $D_i$ derivatives. Since $D_i :=
E^a_i D_a$, the notion does not depend on coordinates. Rather, it is
tied directly to the physical triads and the covariant derivatives
compatible with the metric. Then, the idea behind the conjecture is
that, as one approaches the singularity, the spatial derivatives
$D_i C_{jk},\, D_iP_{jk}$ of the basic fields should become
negligible compared to the basic fields themselves (in particular)
because of the $\sqrt{q}$ multiplier in the definition of $E^a_i$.
An immediate consequence is that the antisymmetric part of $C_{ij}$
is negligible \cite{ahs2}, a fact that we will repeatedly use below.

Thus, our formulation of the BKL conjecture is that, as one
approaches the singularity, solutions to the Einstein's equations
(2.3)--(2.8) are well approximated by solutions to the truncated
system of equations obtained by ignoring the spatial derivatives.
The truncated constraints are purely algebraic equations:
\ba \label{S2} S_{(T)} &:=&  C_{ij} C^{ji} - \f{1}{2} C^2 +
P_{ij} P^{ji} - \f{1}{2} P^2 \approx 0\\
\label{V2} V_i^{(T)} &:=& \epsilon_{ijk} (2P^{jl}C_l{}^k - C P^{jk})
 \approx 0\\
\label{G2} G^{k}_{(T)} &:=& \epsilon^{ijk} P_{ji} \approx 0\, ,
\ea
while the truncated evolution equations are ordinary differential
equations:
\be \label{Cdot2}\dot{C}_{ij} =   {N}\, [2 C_{k(i} P^k{}_{j))} - P
C_{ij}] \quad {\rm and} \quad \dot{P}_{ij} = N [- 2 C_{ik} C^k{}_j +
C C_{ij}]\, . \ee
Note that operation of truncation and hence the final truncated
system depends crucially on one's choice of basic variables and
notions of space and time derivatives. (For example, if we had
used triads rather than $C_{ij}$ as basic variables, we would have
been led to set $C_{ij}$ to zero in the truncation procedure, and
truncation would have led us just to Bianchi I equations. The
resulting BKL conjecture would have been manifestly false.)

How does this formulation compare with that of UEWE \cite{cu,dg1}?
In that framework, one divides the geometrical fields by the trace
of the extrinsic curvature $K$ (which is expected to diverge at the
singularity) while here we multiply them by the volume element
$\sqrt{q}$ (which is expected to go to zero). There, the (scalar)
lapse $\b{N}$ is such that $\b{N}K$ admits a limit $\ub{N}$.
Therefore $\b{N}$ goes to zero and, as in our case, the singularity
lies at $t=-\infty$. The key scale invariant variables $(N_{ij},
\Sigma_{ij})$ which are expected to be well behaved at the
singularity are related to our $(C_{ij}, K_{ij})$ via
\ba N_{ij} = 6 P^{-1}\, C_{ij}\quad &{\rm and}& \quad \Sigma_{ij}
= -6 P^{-1}P_{(ij)} + 2\delta_{ij}\, , \quad {\rm or}, \\
C_{ij} = - \f{K}{3} \sqrt{q} N_{ij} \quad &{\rm and}& \quad P_{(ij)}
= \f{K\sqrt{q}}{3}\, (\Sigma_{ij} - 2\delta_{ij})\, . \ea
Similarly the two sets of lapse fields are related simply by: $N =
\b{N}K\sqrt{q}$. Thus, although the motivations and the starting
points of the two frameworks are quite different, the basic
variables are closely related. From the viewpoint of differential
equations, the two reduced systems would in essence be equivalent if
$K\sqrt{q}$ admits a finite, nowhere vanishing limit at the
singularity. This condition holds for Bianchi I models (and also
Bianchi II which describe the transitions between Bianchi I epochs).
An advantage of the $(C_{ij}, P_{ij})$ framework is that it is
better adapted for non-perturbative quantization since it comes from
the Hamiltonian framework underlying loop quantum gravity.

\textbf{Remarks:}\\
i) In our formulation we have allowed a large class of foliations.
However, a closer examination from the standpoint of differential
equations may well lead to further restrictions. The inverse mean
curvature foliations commonly used in conjunction with the UEWE
framework appear to be well-suited also for
our framework.\\
ii) As a rule of thumb one often says that scale invariant scalars
(such as $N_{ij}$ and $\Sigma_{ij}$) should have well-defined limits
at the singularity. Those considerations can be extended to density
weighted quantities used in this paper. The rule of thumb then is
that a density of weight $n$ has well-defined limit if it has
scaling dimension $2n$. Since $C_{ij}$ and $P_{ij}$ have density
weight 1 and scaling dimension $2$, they should have a well
defined limit at the singularity.\\
iii) Consider the subspace of the full phase space on which $D_i
C_{jk} =0$ and $D_i P_{jk} =0$ and demand that the lapse $N$ satisfy
$D_iN=0$. Then, a non-trivial result is that the Hamiltonian vector
field of \emph{full general relativity} is tangential to this
sub-space; conditions $D_i C_{jk} =0$ and $D_i P_{jk} =0$ are
preserved by the full evolution equations
(\ref{Cdot1})--({\ref{Ddot1}).

\section{The BKL truncated Hamiltonian system}
\label{s4}

Let us now explore the BKL truncated system we were led to in
section \ref{s3} by focusing on the pair $(C_{ij}, P_{ij})$. We
already know that $C_{ij}$ is symmetric due to the BKL truncation.
The Gauss constraint implies that $P_{ij}$ is symmetric. Therefore,
the basic variables of the truncated theory are pairs of
\emph{symmetric} fields $(C_{ij}, P_{ij})$ on $M$.

Upon truncation, the full symplectic structure of general
relativity yields the following Poisson brackets between these
fields:
\ba \{C_{ij}(x),\, C_{kl}(y) \}_{T}&=&0, \quad {\rm and} \quad
\{P_{ij}(x),\, P_{kl}(y)\}_{T} = 0\, ,\nonumber\\
\{P_{ij}(x),\, C_{kl}(y) \}_{T} &=& (C_{k(j}\delta_{i)l} +
C_{l(j}\delta_{i)k})(y)\,\, \delta^3(x,y)\, .\ea
These are subject to the truncated scalar and vector constraints
(\ref{S2}) and (\ref{V2}). This system of constraints is of first
class. As before, for simplicity, let us set the shift to zero and
obtain the evolution equations by taking Poisson brackets with the
scalar constraint. The result is:
\be \dot{C}_{ij} =  N\, [2C_{k(i} P_{j)}{}^k - P C_{ij}], \quad
{\rm and} \quad \dot{P}_{ij} = - N\,[ 2C_{ki} C_{j}{}^k - C
C_{ij}]\, .\ee
This is the Hamiltonian flow generated on the truncated phase space
by the truncated constraints. It agrees with the evolution equations
(\ref{Cdot2}) of section \ref{s3} obtained by first using the full
constraints on the full phase space and then truncating the full
evolution equations. Thus, the truncation procedure is in complete
harmony with the Hamiltonian framework.

To explore the structure of the truncated theory, it is convenient
to solve and gauge fix the vector constraint. Since our fields
$C_{ij}$ and $P_{ij}$ are now symmetric, the vector constraint
(\ref{V2}) implies that, regarded as matrices, they commute,
whence we can simultaneously diagonalize them. Finally, we can
gauge-fix this constraint by demanding that $C_{ij}$ and $P_{ij}$
be diagonal. Then we arrive at the following description of the
truncated phase space. It is coordinatized by the diagonal
elements $C_I, P_I$ of $C_{ij}$ and $P_{ij}$ (with $(I=1,2,3)$).
Their Poisson brackets are given by:
\be \{C_I(x), \, C_J(y)\}_T = 0, \quad \{P_I(x),\, P_J(y)\}_T =0,
\quad \{C_I(x), P_I(y)\}_T = -2 \delta_{IJ} C_J \delta(x,y)\, .\ee
There is a single (scalar) constraint:
\be \label{S3} S_{(T)}(x) := \f{1}{2} \big( \underset{I}{\Sigma} \,
C_I(x) \big)^2 - \underset{I}{\Sigma}\, C_I^2(x) + \f{1}{2}
\big(\underset{I}{\Sigma}\,\, P_I(x) \big)^2 -
\underset{I}{\Sigma}\, P_I^2(x) \approx 0\,. \ee
Thus, as expected, the truncated Hamiltonian system is
ultra-local; dynamics at each spatial point is insensitive to what
is happening elsewhere. Equations of motion are given by:
\be \label {evo} \dot{C}_I(x) = - N C_I(x)\,\,
\big(\underset{J}{\Sigma} P_J(x) - 2P_I(x)\big)\, ,\quad
\dot{P}_I(x) = N C_I(x)\,\, \big(\underset{J}{\Sigma} C_J(x) -
2C_I(x)\big)\, .\ee
Note that, although the constraint is symmetric under interchange of
$C_I$ and $P_I$, there is a basic asymmetry in the symplectic
structure which descends to the equations of motion. Dynamics leaves
the sector of the phase space with $C_I \ge 0$ invariant. Let us
focus on this sector and set $Q_I = - \ln C_I/2$. Then, $(Q_I, P_I)$
are canonically conjugate. Rewriting the constraint (\ref{S3}) in
terms of $Q_I$, one immediately sees that what we have is particle
dynamics in exponential potentials. By making a rigid rotation in
the truncated phase space, (\ref{S3}) can be brought to the familiar
Misner form \cite{bb} $S_{(T)} = -\b{P}_0^2 +\b{P}_+^2 + \b{P}_-^2 +
e^{-(4/\sqrt{3}) \b{Q}_0}\, V(\b{Q}_\pm)$. However, $\b{Q}_0,
\b{Q}_\pm$ are components of a connection while the Misner variables
refer to components of the 3-metric.

\begin{figure}[tbh!]
  \begin{center}
    \includegraphics[width=7in,height=3in,angle=0]{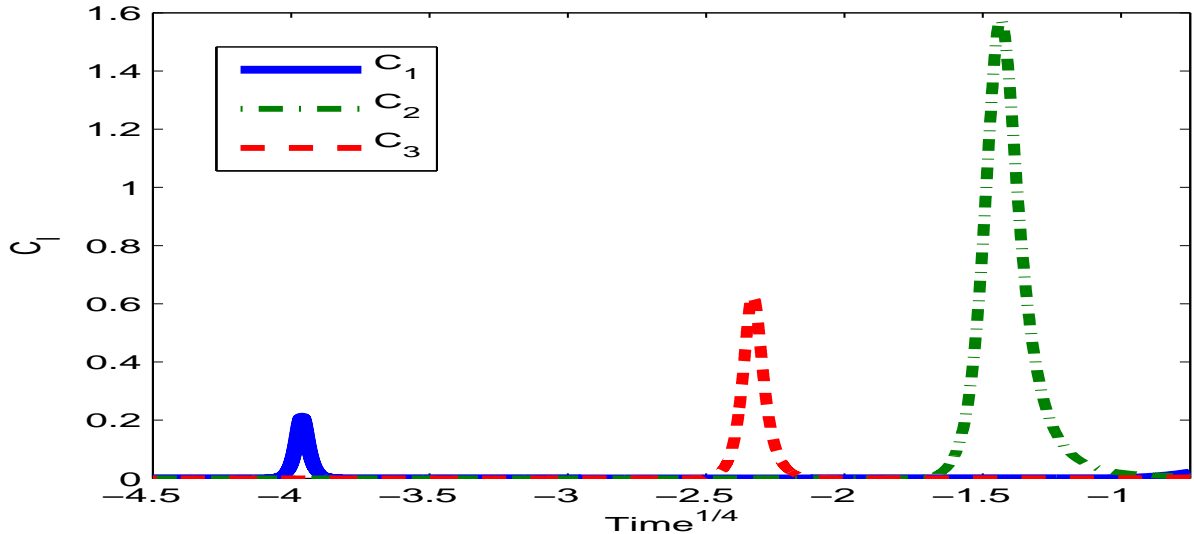}
\caption{Past evolution of the three $C_I$ at a fixed point $x$,
given by Eq (\ref{evo}). As time decreases, $C_2$ starts growing
rapidly near $t=-1$,
reaches a peak, and then goes to zero rapidly. Then near $t= -23$,
$C_3$ follows the same pattern and finally $C_1$ does the same
near $t= -230$. Their profiles are well described Eq (\ref{sol1}).
This growth and decay repeat ceaselessly as the singularity is
approached at $t=-\infty$. In this simulation, the initial data at
$t=0$ was $P_1= -0.4742, P_2= -3.3586, P_3= -1.3096; C_1= 0.0746,
C_2= 0.0040, C_3= 0.0070$.}
    \label{fig1}
  \end{center}
\end{figure}

\begin{figure}[tbh!]
  \begin{center}
    \includegraphics[width=7in,height=3in,angle=0]{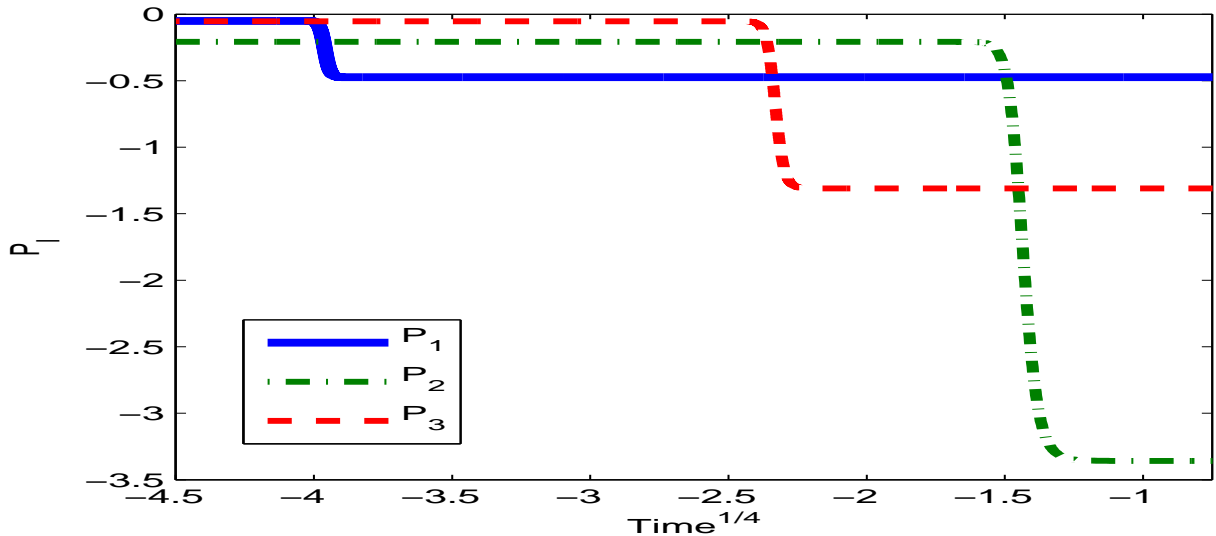}
\caption{Past evolution of the three $P_I$ at a fixed point $x$
(in the same simulation as in Figure \ref{fig1}). Their profiles
are well described by Eq (\ref{sol2}). Only one of the three $P_I$
changes at a time, causing the solution to make a transition from
one Bianchi I metric to another. The jump in each $P_I$ is
correlated in time with that in the corresponding $C_I$.}
    \label{fig2}
  \end{center}
\end{figure}

Note that if $C_I(x)$ vanish initially at a point $x$ in $M$, they
remain zero throughout the evolution and $P_I(x)$ remain constant.
Then the dynamics of fields at that point $x$ is that of the Bianchi
I model.%
\footnote{Note that since $M$ is allowed to have any compact
topology, it may not admit a global Bianchi I solution.}
(The $P_I$ are linear combinations of the parameters $p_i$ normally
used to characterize the Bianchi I metric.) Dynamics of generic
initial data is much more complicated than what one might expect
from the deceptively simple form of the evolution equations
(\ref{evo}). To get a handle on the Bianchi II transitions, one can
linearize (\ref{evo}) about a Bianchi I solution. Since we have
assumed that the universe contracts as $t$ decreases, it follows
that all three $P_I$ of the Bianchi I solution are negative. One
finds that as one evolves backward in time two of the the three
modes $(C_I,P_I)$ are stable and one, with largest $|P_I(x)|$, is
unstable. Let us suppose that $|P_1(x)|$ is the largest. Then the
exact evolution of the stable modes 2 and 3 is well tracked by the
linearized equations which imply $C_2(x,t) \approx 0 \approx
C_3(x,t)$ and $P_2(x,t) \approx P_2(x,t_0), P_3(x,t) \approx
P_3(x,t_0)$. However, for the unstable mode we have to use the full
equations (\ref{evo}). Setting $C_2 = C_3 =0$ one obtains the
solution:
\ba \label{sol1} C_1(x,t) &=& \pm 2 ( \sqrt{P_2P_3}\,\, {\rm
sech}\, (2\sqrt{P_2P_3}N\, (t-\tau)),\, {\rm and} \\
\label{sol2} P_1(x,t) &=& P_2 + P_3 + 2\sqrt{P_2P_3}\, \tanh
(2\sqrt{P_2P_3} N\,(t-\tau)) \ea
for some constant $\tau$, where the sign in the first equation is
plus (minus) if $C_1(x,t_0)$ is positive (negative). Thus, only one
pair, namely $(C_1(x,t),P_1(x,t))$, evolves non-trivially. The form
of solutions (\ref{sol1}) implies that, as we evolve backward in
time, there is a quick transition in which $|C_1(x)|$ first
increases, reaching a maximum value at $t=\tau$, and is then driven
to zero, while $P_1(x)$ is mapped to $P_2(x,t_0)+ P_3(x, t_0) + 2
\sqrt{P_2(x,t_0) P_3(x,t_0)}$. Simple algebra shows that this
transition from one Bianchi I solution to another is given precisely
by the well-known $u$-map. This derivation of the $u$-map does not
require one to approximate the exponential potential in (\ref{S3})
by a rigid, perfectly reflecting wall. Also, our parametrization of
the Bianchi-I solutions by $P_I$ (rather than by the standard $p_i$)
makes it easier to follow the Bianchi II transitions both
analytically and numerically because in any transition \emph{only
one} of the three $P_I$ changes (while all three $p_i$ change).
Figures \ref{fig1} and and \ref{fig2} show these transitions in a
numerical solution to the ODEs (\ref{evo}). Here, an iterative
Picard algorithm was used with an adaptive time step, chosen to
ensure conservation of the scalar constraint. Such transitions in
the truncated model mimic the situation in the full theory
surprisingly well. (Compare, for example, our figures \ref{fig1} and
\ref{fig2} with figures 5 and 4 in the second paper of \cite{dg1}).

So far we focused on the dynamics at a single spatial point $x$.
Let us now consider a 2-dimensional surface in the physical space
on which one of the $C_I$, say $C_1$, vanishes and $C_2,C_3$ are
small. Fix a point $x$ on this surface. Then, $C_1$ is positive to
one side of the 2-surface and negative to the other. At a point
$y$ in a neighborhood of $x$, as $t$ decreases, $C_1(y)$ first
increases rapidly if it is initially positive and decreases
rapidly if it is initially negative (following the ${\rm
sech}(t-\tau)$ profile of (\ref{sol1})). This produces steep
gradients at $x$ for some time, which appear as spikes. However,
as $t$ decreases further, $|C_1|$ goes to zero rapidly on both
sides of the 2-surface for $t<\tau$. So the spike in $C_1$ becomes
dilute and disappears (but it will reoccur and disappear again
repeatedly on further backward evolution). $P_1$ on the other hand
remains constant at $x$ but decreases on either side of the
2-surface. As $t$ decreases, the gradient of $P_1$ keeps
increasing and the spike sharpens. These transitions and spikes
have been observed in numerous simulations of the truncated system
\cite{hs}. However, because the spatial gradients become large at
spikes, the truncated system becomes a poor approximation and one
has to consider the full system (as, e.g., in \cite{wcl}).

The analytical arguments given above explain all the qualitative
features of the truncated dynamics if the initial phase space point
lies in a neighborhood of a Bianchi I fixed point. But we do not
have an analytical understanding of what happens away from this
neighborhood. Numerical simulations show that even if one starts far
away from pairs $(C_I, P_I)$ corresponding to Bianchi I solutions,
dynamics drives the system quickly to the Bianchi I sub-space. It
would be very useful to have an analytical derivation of this
phenomenon for our system (\ref{evo}) without having to make further
approximations.

\section{Discussion}
\label{s5}

We began with the Hamiltonian formulation of general relativity
underlying loop quantum gravity where the basic fields are spatial
triads $E^a_i$ with density weight 1, spin connections $\Gamma_a^i$
they determine, and extrinsic curvatures $K_a^i$. Based on the
examples that have been studied analytically and numerically, the
general expectation is that the determinant $q$ of the spatial
metric $q_{ab}$ would become degenerate and the trace $K$ of the
extrinsic curvature would diverge at space-like singularities. One
can therefore hope to obtain fields which remain well-defined at the
singularity either by multiplying natural geometric fields by
suitable powers of $q$ or dividing them by suitable powers of $K$.
In a commonly used UEWE framework \cite{cu,dg1}, one chooses to
divide by $K$. The resulting fields satisfy differential equations
with desirable properties. However, because of the presence of
$K^{-1}$, in quantum theory it is difficult to introduce operators
corresponding to the new fields.

We adopted the complementary strategy of multiplying geometrical
fields by $\sqrt{q}$. In the Hamiltonian formulation with which we
began, the basic field $E^a_i$ is already obtained by multiplying
the orthonormal triad by $\sqrt{q}$. One would therefore expect it
to vanish at the singularity and this expectation is borne out in
examples. Consequently, $E^a_i$ also provides a natural avenue to
construct additional fields needed in the BKL conjecture. Our
variables $C_i{}^j$ and $P_i{}^j$ were obtained (modulo trace terms)
simply by contacting the spatial indices of $\Gamma_a^j$ and $K_a^j$
by $E^a_i$. Furthermore, because $E^a_i$ vanishes in the limit, the
operator $D_i:= E^a_i D_a$ provides a convenient tool to express the
notion of \emph{`spatial derivatives which are to become
subdominant'} near the singularity. The main expectation is that
asymptotically $D_iC_{jk}$ and $D_i P_{jk}$ would become
`negligible' relative to $C_{jk}$ and $P_{jk}$. Now, in exact
general relativity, time derivatives of $C_{ij}$ and $P_{ij}$ can be
expressed in terms of their $D_i$ derivatives, purely algebraic (and
at most quadratic) combinations of $C_{ij}$ and $P_{ij}$, the lapse
$N$ and its $D_i$ derivatives (see (\ref{S1})--(\ref{Ddot1})).
Therefore, if in the limit the $D_i$ derivatives of the basic fields
become negligible compared to the fields themselves, we are
naturally led to conclude that time derivatives would dominate the
spatial derivatives. This chain of argument led to our formulation
of the BKL conjecture.

This rather simple idea depends on the fact that the structure of
Einstein's equations has an interesting feature: as saw in section
\ref{s2}, once the triplet $C_{ij}, P_{ij}, D_i$ is constructed from
the triad $E^a_i$ and the extrinsic curvature $K_a^i$ on an
\emph{initial slice}, the constraint and evolution equations can be
expressed entirely in terms of the triplet. Given a solution to
these equations, the spatial triad $E^a_i$ (and hence the metric
$q_{ab}$) can be recovered at the end simply by solving a total
differential equation, (\ref{Edot}). This is a surprising and
potentially deep property of Einstein's equation. It provided key
motivation for our formulation of the BKL conjecture and could well
capture the essential reason behind the BKL behavior observed in
examples and numerical simulations. Therefore, an appropriate
quantization of the truncated system, e.g., a la loop quantum
cosmology, could go a long way toward understanding the fate of
generic space-like singularities in quantum gravity.

Since the framework is developed systematically from a Hamiltonian
theory, the BKL truncation naturally led to a truncated phase space.
The specific truncation used has an important property: The
truncated constraint and evolution equations on the truncated phase
space coincide with the truncation of full equations on the full
phase space. On the truncated phase space we could solve and
gauge-fix the Gauss and vector constraints to obtain a simple
Hamiltonian system. Solutions to this system have been explored both
analytically \cite{ahs2} and numerically\cite{hs}. They exhibit the
Bianchi I behavior, the Bianchi II transitions and spikes as in the
analysis of symmetry reduced models \cite{bb} and numerical
investigations of full general relativity \cite{dg1}.

Finally, in the main text we have restricted ourselves to vacuum
equations. The addition of a massless scalar field is
straightforward because the Hamiltonian framework with which we
began can be easily extended to accommodate a scalar field
\cite{art}. The main features that are generally expected from the
analysis of Andersson and Rendall \cite{ar} are reflected in the
resulting truncated system. Thus, if the energy density in the
scalar field is small, one again has Bianchi II transitions and
spikes. However, once the energy density exceeds a critical value,
these disappear and the asymptotic dynamics at any spatial point is
described just by the Bianchi model with a scalar field without any
transitions.

Derivations of analytical results and numerical simulations will
appear in detailed papers \cite{ahs2,hs}.
\bigskip

\textbf{Acknowledgments:} We are would like to thank Woei Chet Lim,
Alan Rendall, Claes Uggla, Yuxi Zheng and especially David Garfinkle
for discussions. This work was supported in part by the NSF grants
PHY0456913 and PHY0854743, the Eberly research funds of Penn State
and a Frymoyer Fellowship to DS.

\end{document}